# Physics-based Analytical Modeling of CMOS Latchup

Gennady I. Zebrev*

**Abstract—Analysis of the steady-state Kirchhoff equations within the framework of a new physics-based equivalent circuit provides explicit expressions for the holding voltage and current for the resistive mode in bi-stable CMOS structures. The resulting expressions are functions of the physical parameters (temperature, doping level of the substrate, material band gap), which makes it possible to evaluate the influence of technological factors on the latchup parameters. In particular, an explicit formula was obtained for the critical charge for single event latchup as a function of process parameters and temperature.**

## I. Introduction

It is well-known that external non-stationary impact can generate parasitic low-impedance path for the passage of current between the ground and power contacts in CMOS integrated circuits, causing the so-called latchup (parasitic thyristor) looking like a jump in the IC's supply current of hundreds of milliamps or even several amperes[1, 2]. Due to its potentially destructive nature, this process is one of the biggest challenges in modern digital electronics. One of the common mechanisms of external influence is local ionization by high-energy heavy ions in outer space (Single Event Latchup, SEL). [3,4, 5,6,7]. It is known that the vulnerable sensitive area of a CMOS structure is the reverse-biased n-well/ p-well junction on which the entire supply voltage $V_{DD}$ drops in statics [8]. This pn-junction, when exposed to an ionizing particle, plays the role of a current source (generator), leading to a resistive voltage drop in the n- and p-wells. This in turn causes a forward bias of the n- and p-MOSFET's sources with a positive feedback effect on the primary ionization current. As a result, a stationary current is established in the npnp structure, at which the supply voltage $V_{DD}$ is divided between the resistances of the n- and p-wells. The latchup simulation is usually carried out at the circuit level using the so-called "two-transistor equivalent circuit", in which the middle pn-junction is represented as collector junctions of two virtual NPN and PNP bipolar transistors simultaneously[9, 10]. The main drawback of the two-transistor model is that it is based on description in terms of parameters of abstract schematic components not tied to parameters of real physical structures. In particular, it is impossible to unambiguously relate the gain parameters of virtual transistors to real dimensions, geometry, temperature and material characteristics. For this reason, circuit calculations within the framework of the two-transistor model are mostly illustrative in nature and cannot be used for quantitative characterization of specific circuits.

We present in this work an analytical physics-based model based on the analysis of stationary Kirchhoff's equations for a more physical equivalent circuit of an npnp structure, allowing us to explicitly obtain analytical expressions expressed in terms of the physical parameters of the structure. The mechanisms of latchup transient initiation will not be discussed here.

* Gennady I. Zebrev is with the Department of Micro- and Nanoelectronics of National Research Nuclear University MEPHI, Moscow, Russia (e-mail: gizebrev@mephi.ru).

## II. Physics-based equivalent circuit and Kirchhoff's equations for CMOS npnp structure

Fig. 1 shows a simplified, but still physically meaningful equivalent circuit with three diodes (n+ and p+ sources of complementary MOSFETs and the middle PW/NW junction).

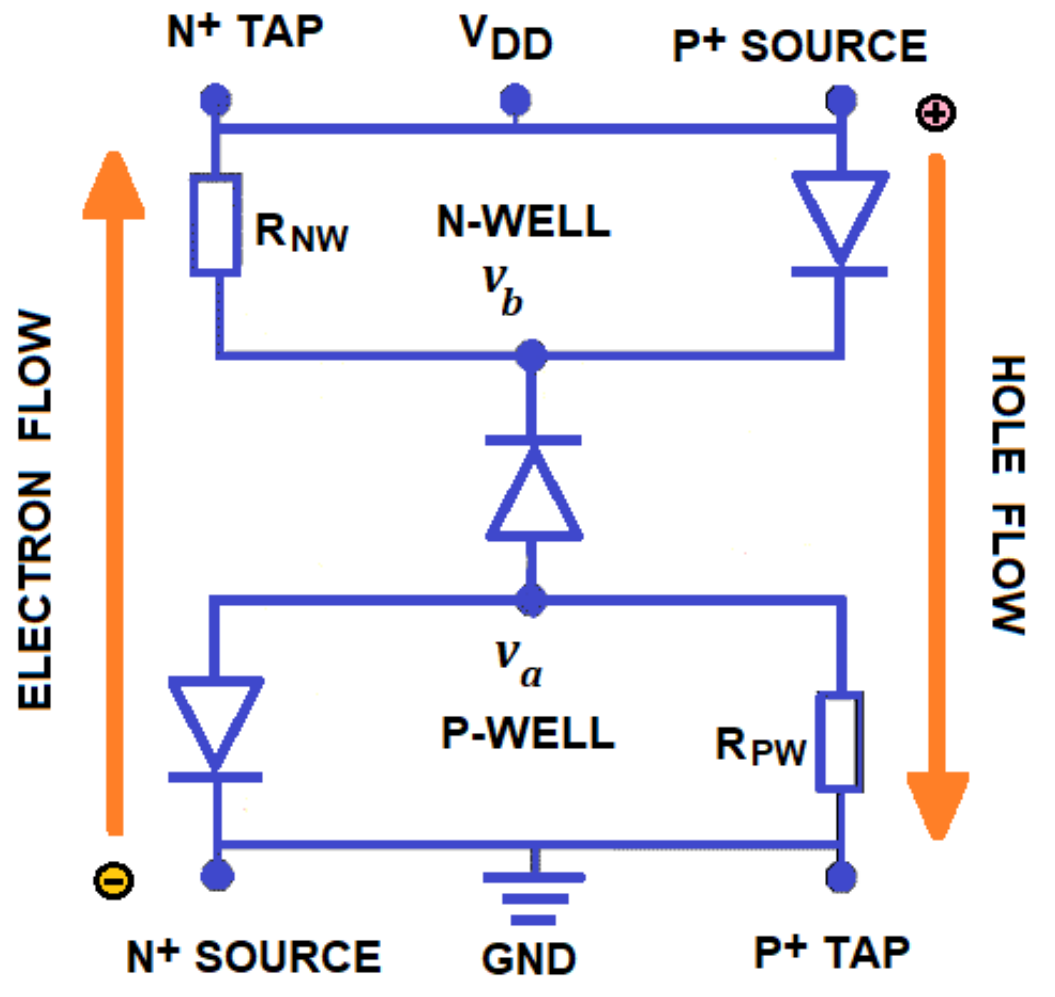

Fig. 1 Three-diode equivalent circuit for physics-based analysis of latchup effects.

Note that the role of diodes in this equivalent circuit is played by the pn junctions of the sources and the middle pn junction between the pockets, which have real physical dimensions and characteristics. From a physical point of view, the current in resistive latch mode consists of electron and hole streams flowing towards each other from the n- and p-MOSFET's sources to the supply and ground contacts. The flows of electrons and holes are not independent and are connected to each other through a positive feedback mechanism. For example, the flow of electrons injected from the $n^+$ source into the p region, due to the ohmic voltage drop in the n-region, opens the $p^+$-source injecting holes, which in turn maintain the steady-state injection of electrons from the $n^+$ source. Immediately after injection from forwardly biased sources, these minority carrier streams propagate in a diffusion manner, and then, after passing through the middle NW/PW junction, they turn into majority carriers and propagate in a drift manner. Neglecting recombination, the diffusion and drift currents in the different wells will be the same both for of electrons and holes. Then the Kirchhoff's equations set for equivalent circuit in Fig. 1 can be written as follows

$$\frac{v_a}{R_p} = I_{p0}\left(\exp\left(\frac{V_{DD}-v_b}{\varphi_T}\right)-1\right), \quad (1a)$$

$$\frac{V_{DD}-v_b}{R_n} = I_{n0}\left(\exp\left(\frac{v_a}{\varphi_T}\right)-1\right) \quad (1b)$$

where $V_{DD}$ is the supply voltage, $\varphi_T = k_B T/q$ is the thermal potential, $v_a$ and $v_a$ are the well potentials, $I_{n0}$ and $I_{p0}$ are the source saturation currents, $R_n$ and $R_p$ are effective series resistances of low-impedance paths in n- and p-wells. These equations can be also interpreted as equalities of diffusion and drift currents for holes and electrons in different wells. It is useful to introduce a dimensionless order parameter $\beta \equiv 1-(v_b - v_a)/V_{DD}$. Then the total voltage drop across the npnp structure $V_{DD} = v_a + (V_{DD}-v_b) + V_R$ [11] can be written as $V_{DD} = V_R + \beta V_{DD}$, where $V_R$ is the reverse bias on the middle pn-junction. The trivial blocking solution of (1) ($v_a = 0$, $v_b = V_{DD}$) corresponds to uniform quasi-Fermi levels in n- and p-wells. In this mode, $\beta = 0$, $V_R = V_{DD}$, and the current in the structure is formally zero. It should be noted that in real structures the current is never strictly zero due to the reverse current of the middle pn junction and, more importantly, due to possible leakage under the STI oxide [12]. In addition, such inter-transistor leakage currents are global and for modern process technologies (< 100 nm) can be significant (tens and hundreds of mA in the IC power circuit), which makes it difficult to characterize the local latchup currents.

## *A. Symmetric case approximation*

Equations (1) can be rewritten in an equivalent form

$$v_a = \varphi_T\left(e^{\frac{V_{DD}-v_b-V_p^*}{\varphi_T}}-1\right), \quad (2a)$$

$$V_{DD} - v_b = \varphi_T\left(e^{\frac{v_a-V_n^*}{\varphi_T}}-1\right), \quad (2b)$$

where

$$V_n^* \equiv \varphi_T \ln\left(\frac{\varphi_T}{v_{n0}}\right), \quad V_p^* \equiv \varphi_T \ln\left(\frac{\varphi_T}{v_{p0}}\right), \quad (3)$$

are the "knee" voltages of the source pn-junctions (~0.5-0.6 V), and $v_{p0} \equiv I_{p0}R_p$, $v_{n0} \equiv I_{n0}R_n$.

For the symmetric CMOS structure, $v_{n0} = v_{p0} \equiv I_0 R_0$ ($V_n^* = V_p^* \equiv V^*$), we have a solvability condition $v_a = V_{DD} - v_b$ which can be interpreted as the equality of the splitting of nonequilibrium quasi-Fermi levels in two wells (Fig. 2). Neglecting in (2) the terms $v_{n0}$ and $v_{p0}$, one can find a solution satisfying the conditions $v_a + v_b = V_{DD}$ and $v_b - v_a = V_R$

$$v_{a,b} \cong \frac{V_{DD} \mp V_R}{2} \quad (4)$$

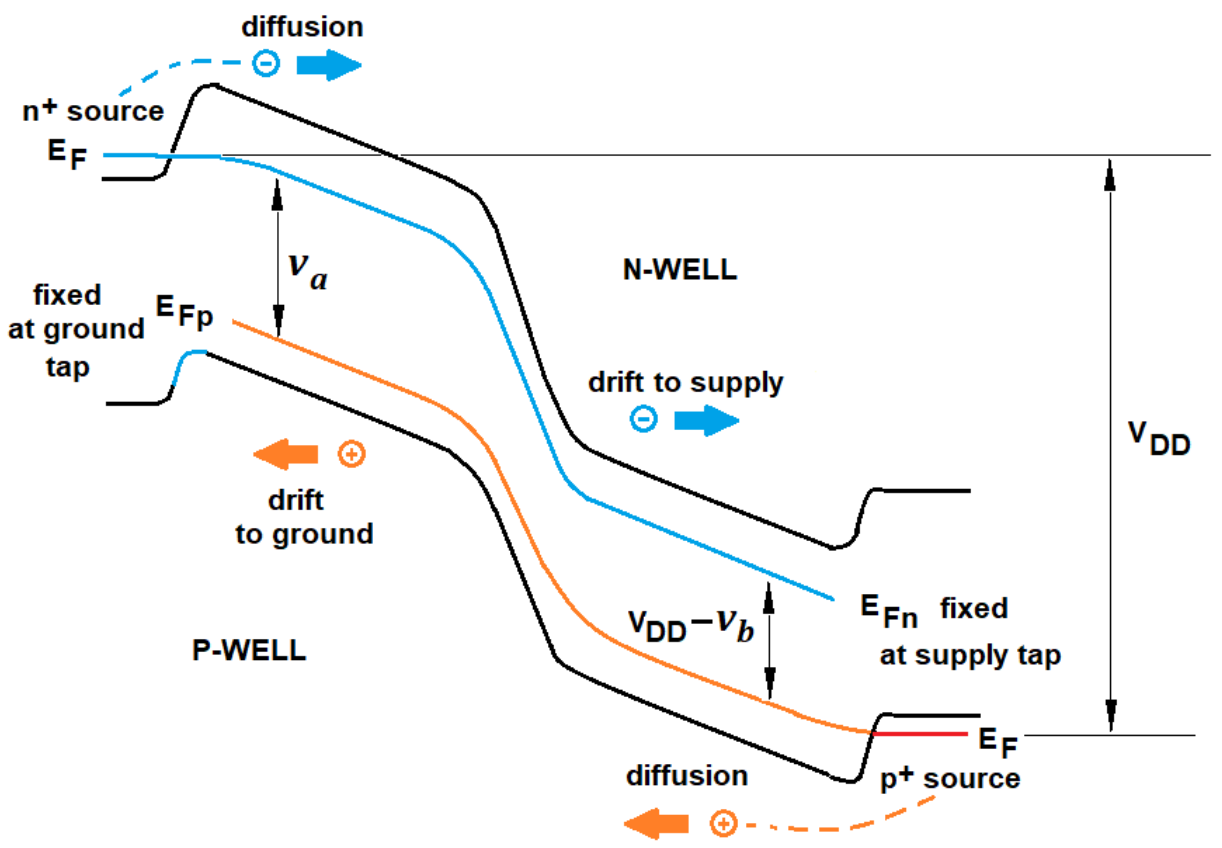

Fig. 2 Band diagram of the npnp structure in a resistive mode.

where the reverse bias of the middle pn-junction is expressed as

$$V_R = \left(V_{DD}^2 - 4\varphi_T^2 e^{\frac{V_{DD}-2V^*}{\varphi_T}}\right)^{1/2}. \quad (5)$$

Fig. 3 illustrates solutions (4-5).

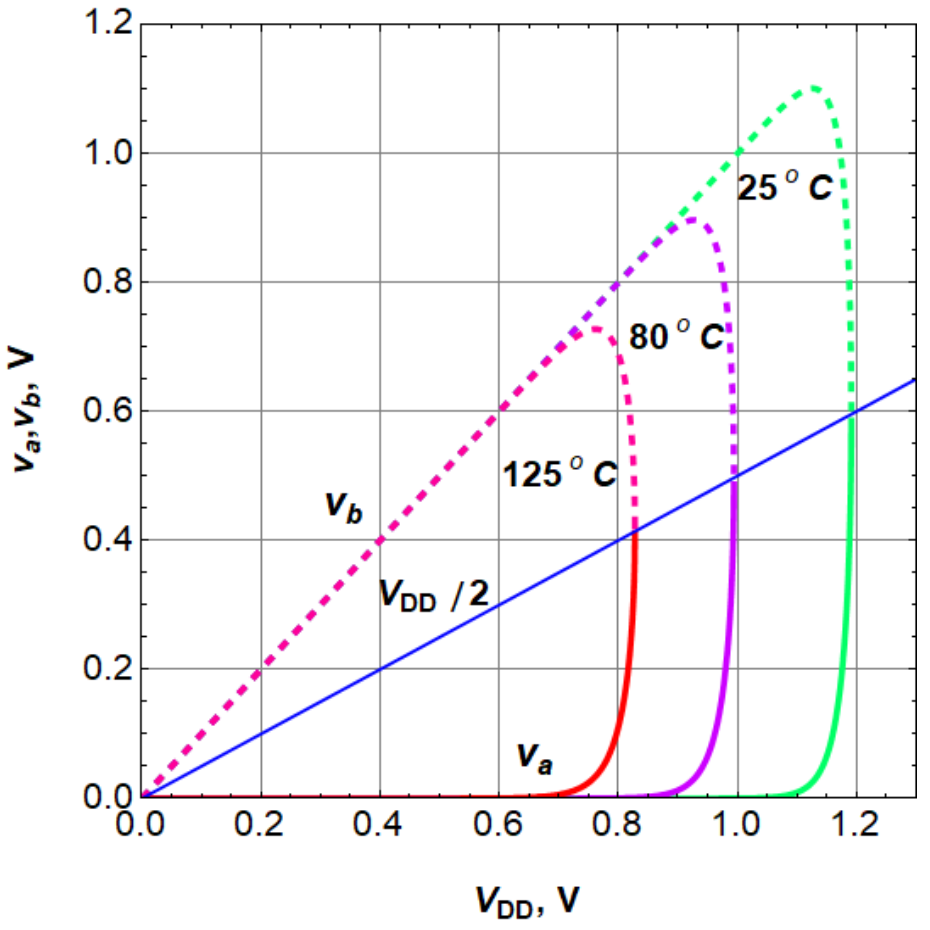

Fig. 3 Potentials $v_a$ and $v_b$ calculated with (4-5) at three different temperatures.

A special case $v_a = v_b = V_{DD}/2$ corresponds to an unbiased middle junction regime $V_R = 0$ which realized under the condition $V_{DD}/2 = \varphi_T \exp\left[\left(V_{DD} - 2V^*\right)/2\varphi_T\right]$. This equation has an exact solution

$$V_H = -2\varphi_T W_{-1}\left[-\frac{v_0}{\varphi_T}\right] = -2\varphi_T W_{-1}\left[-\exp\left(-\frac{V^*}{\varphi_T}\right)\right] \quad (6)$$

where $W_{-1}(x)$ is a non-principal branch of the Lambert function defined as a solution of $W\left(xe^x\right) = x$ [13]. It is important here that the Lambert function is multivalued for negative arguments and the solution (6) is its non-principal branch. Physically $V_H$ corresponds to the intrinsic holding voltage, i.e., the minimum voltage $V_{DD}$ at which resistive current mode in the structure is possible. It is easy to obtain a simpler and more transparent approximate expression, convenient for practical applications

$$V_H \cong 2V^* + 2\varphi_T \ln\frac{V^*}{\varphi_T} = 2\varphi_T \ln\frac{V^*}{I_0 R_0}. \quad (7)$$

### *B. General case of asymmetric structure ($v_{n0} \neq v_{p0}$)*

Generally, we can derive the exact solutions of (2)

$$v_a = \varphi_T W\left[\frac{v_{p0}}{\varphi_T} e^{\frac{\beta V_{DD} + v_{p0}}{\varphi_T}}\right] - v_{p0}, \quad (8a)$$

$$v_b = V_{DD} - \varphi_T W\left[\frac{v_{n0}}{\varphi_T} e^{\frac{\beta V_{DD} + v_{n0}}{\varphi_T}}\right] + v_{n0}, \quad (8b)$$

identically satisfying to trivial blocking condition ($v_a = 0$, $v_b = V_{DD}$) at $\beta = 0$. The non-trivial solvability condition of (8) $v_b = v_a$ at $\beta = 1$ yields an equation for determining $V_H$:

$$V_H = \varphi_T W\left[\frac{v_{n0}}{\varphi_T} e^{\frac{V_H + v_{n0}}{\varphi_T}}\right] + \varphi_T W\left[\frac{v_{n0}}{\varphi_T} e^{\frac{V_H + v_{n0}}{\varphi_T}}\right] - v_{n0} - v_{p0} \quad (9)$$

Using an asymptotic form of Lambert function $W(x) \cong \ln x - \ln[\ln x]$ at $x > 1$, we get

$$V_H \cong V_n^* + V_p^* + \varphi_T \ln\left(\frac{V_H - V_n^*}{\varphi_T}\right) + \varphi_T \ln\left(\frac{V_H - V_p^*}{\varphi_T}\right) \cong \cong \varphi_T \ln\left(\frac{V_n^*}{v_{n0}}\right) + \varphi_T \ln\left(\frac{V_p^*}{v_{p0}}\right), \quad (10)$$

which is essentially the same as (7) for symmetric case at $I_0 R_0 \equiv \sqrt{v_{n0} v_{p0}}$ and $V^* \equiv \sqrt{V_n^* V_p^*}$. The total current of the structure is a sum of electron and hole components

$$I_D = I_p + I_n = \frac{v_a}{R_p} + \frac{V_{DD} - v_b}{R_n} = = \frac{\varphi_T}{R_p} W\left[e^{\frac{V_{DD} - V_p^* + v_{p0}}{\varphi_T}}\right] + \frac{\varphi_T}{R_n} W\left[e^{\frac{V_{DD} - V_n^* + v_{n0}}{\varphi_T}}\right] - \frac{v_{p0}}{R_p} - \frac{v_{n0}}{R_n}, \quad (11)$$

which is identically equal to zero at $V_{DD} = 0$. When $V_{DD} = V_H$ we have a logarithmic asymptote which yields an analytical expression for holding current

$$I_H \equiv I_D\left(V_{DD} = V_H\right) \cong \frac{\varphi_T}{R_n} \ln\left(\frac{V_n^*}{v_{p0}}\right) + \frac{\varphi_T}{R_p} \ln\left(\frac{V_p^*}{v_{n0}}\right). \quad (12)$$

Essentially, this is the sum of the electron and hole components $I_H = I_{pH} + I_{nH}$ that satisfy the condition $V_H = I_{pH} R_p + I_{nH} R_n$ which is consistent with (10). The structure current in the resistive mode ($V_{DD} > V_H$, $I_D > I_H$), we have in the same approximation

$$I_D \cong I_H + \frac{V_{DD} - V_H}{R_{\parallel}}, \quad (13)$$

where $R_{\parallel} \equiv \left(R_n^{-1} + R_p^{-1}\right)^{-1}$. The extrinsic holding voltage $V_H^{ext}$ in a circuit with an external load resistance $R_L$ can be significantly greater than the intrinsic one $V_H^{ext} \equiv V_H + I_H R_L$.

## III. Validation and applications

In practice, we usually do not have information about the degree of CMOS asymmetry, so for evaluations we will use some effective values. The temperature dependence of the holding voltage $V_H$ in (6-7) is mainly controlled by the dimensionless model parameter

$$\frac{I_0 R_0}{\varphi_T} \cong a\left(\frac{n}{N_W}\right)^2 \propto T^3 e^{-\frac{E_G}{kT}}, \quad (14)$$

where $N_W$ is an effective well doping level, $E_G$ and $n_i$ are the Si bandgap and intrinsic density, *a* is a dimensionless geometric factor that cannot be calculated analytically due to the complex geometry of the system and distributed resistances [14]. Note that the temperature dependences of the diffusion coefficient $D_0$ and mobility in resistance cancel each other out and only the fundamental temperature dependence of the square of the intrinsic concentration remains

$$V_H \cong -2\varphi_T W_{-1}\left[-a\left(n_i/N_W\right)^2\right]. \quad (15)$$

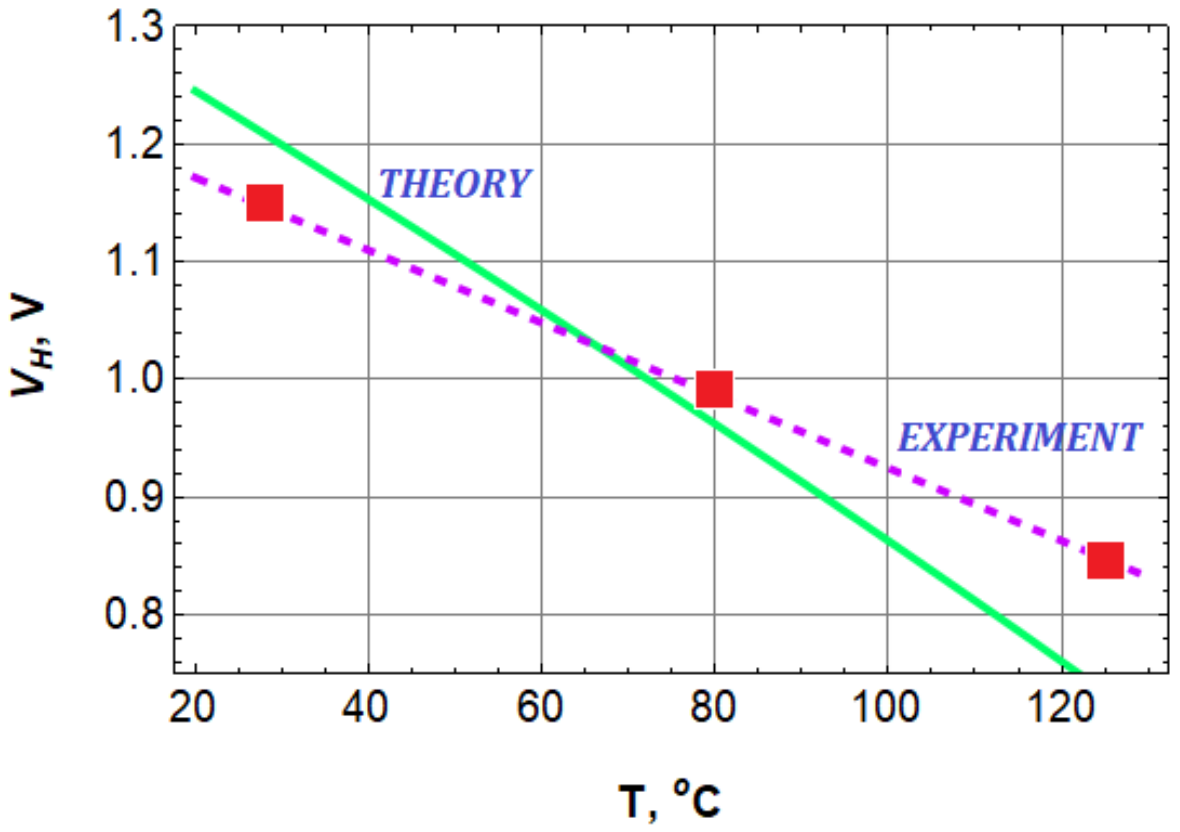

Fig. 4 Comparison of simulation with (15) and experimental data in [15]. Fitting parameters are $a = 40$, $N_W = 2\times10^{15}$ cm$^{-3}$.

Fig. 4 shows a comparison of model simulation and recent experimental data for 7nm FinFETs [15]. Note that $V_H$ depends on the geometric factor $a$ only logarithmically and the uncertainty in the choice of $a$ has very little effect on the result.

The discrepancy between the model and experiment may be due to a systematic error in the experimental determination of $V_H$ associated with masking of the local latchup current by the global leakage current of the entire IC, which has a strong temperature dependence. Interestingly, that TCAD simulations produce a similar systematic discrepancy [16, 17]. Using (7), one can obtain an approximate analytical expression for the holding voltage temperature coefficient

$$\frac{dV_H}{dT} \cong \frac{V_H - 2E_G/q}{T} + 2\alpha_G \,, \qquad (16)$$

where $\alpha_G \equiv dE_G/dT \sim -2\times10^{-4}$ V/K (Si) is the Si bandgap temperature coefficient. Typical value of the first term for silicon at room temperature ~ - 3 mV/K, the contribution of the temperature bandgap change is about ~ 0.4 mV/K. Figs. 5-6 show the holding voltage as functions of temperatures and well doping simulated with (15) at different parameter values. Note that the holding voltage increases with increasing doping level and decreasing temperature, as is the case experimentally.

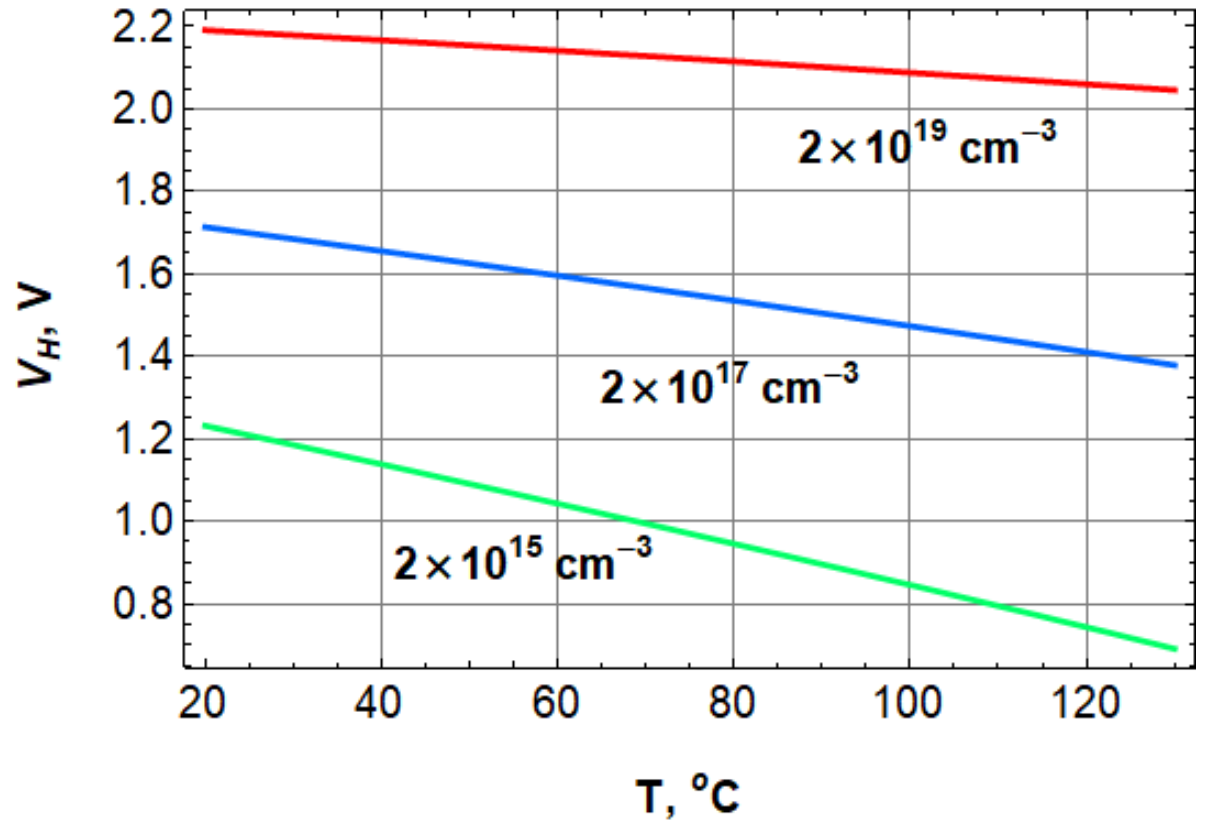

Fig. 5 Temperature dependencies of holding voltage simulated with (15) for different doping levels.

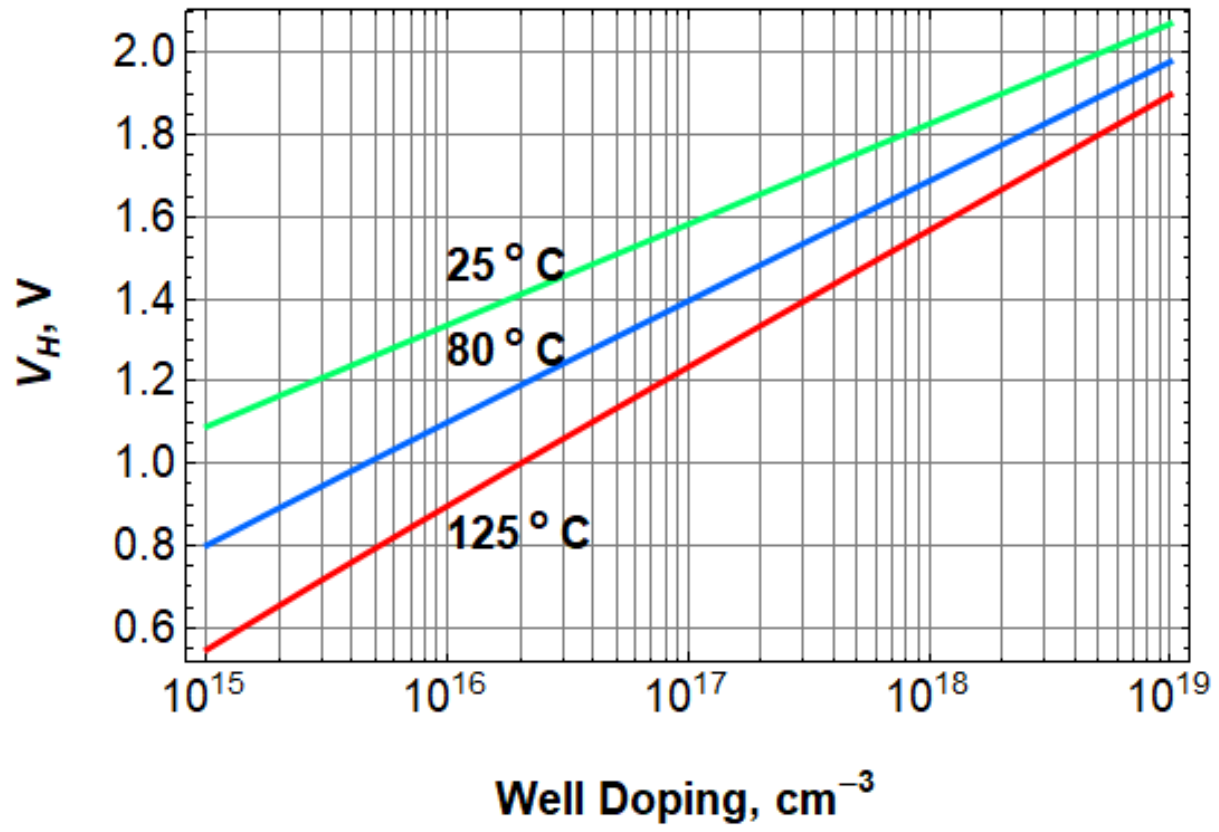

Fig. 6 Holding voltage simulated as functions of well doping at three different temperatures.

Taking into account the possible decrease in effective doping with decreasing temperature due to the freeze-out effects, it can be expected that $V_H(T)$ will have a maximum in the low temperature region. This assumption is completely confirmed experimentally [18, 19] and by TCAD calculations in [20], which could be additional evidence of physical equivalence (at least at a qualitative level) between the analytical model and TCAD simulation.

The results obtained can be useful when modeling the SEL cross sections [21] or, for example, for simulating of the Si CMOS-based single neuron devices [22].

## IV. Single Event Latchup cross section

The tolerance of digital circuits to the impact of single ionizing particles is commonly characterized by the parameter of critical charge $Q_C$. The probability (cross section) of a rare single event (e.g., SEL) can be estimated through the critical charge $Q_C$ as follows [21, 23, 24]

$$\sigma_{SEL} = A_S \exp\left(-\frac{Q_C}{\Delta Q_{coll}}\right), \qquad (17)$$

where $A_S$ is a sensitive area of the CMOS integrated circuit, $\Delta Q_{coll}$ is a charge (for SEL, typically $< Q_C$) released and collected in the sensitive nodes during the passage of an ionizing particle. For heavy ion induced direct ionization we have

$$\Delta Q_{coll} = \frac{\rho_{Si} t_{eff}}{\varepsilon_p} \Lambda \qquad (18)$$

where $\Lambda$ is the linear energy transfer (LET) of a given ion, $\rho_{Si}$ is the silicon mass density, $\varepsilon_p$ is the energy required to create one electron-hole pair (~ 3.6 eV in Si), $t_{eff}$ is the effective length for charge collection at a fixed angle of incidence of the ion.

The critical charge should be generally proportional to the product of the minimum amplitude and the minimum duration of the voltage surge sufficient for the triggering of the single event

$$Q_C = \frac{V_{\min}\tau_{\min}}{R_0} \qquad (19)$$

where $R_0$ is resistance through which the injected charge relaxes. The minimum voltage required for the SEL is the holding voltage $V_H$. The minimum time scale for turning on the SEL is the time of diffusion $\tau_{DIFF}$ of nonequilibrium electrons and holes injected from the FETs' drains to the middle pn-junction (these are time scales on the order of nanoseconds [25]). Following the general formula (19), we write the critical charge for the SEL in the form

$$Q_C = \frac{V_H \tau_{DIFF}}{R_0} \,. \qquad (20)$$

For simplicity, we will use here the simplest estimates

$$\tau_{DIFF} \cong \frac{L_W^{\;2}}{\mu_0 \varphi_T} \,, \qquad R_0 \cong \frac{L_W}{A q \mu_0 N_W} \,, \qquad (21)$$

where $A$ and $L_W$ are characteristic length and cross-sectional area of the wells. Taking into account (15) the formula for the critical charge takes the simple form

$$Q_C \cong \frac{V_H}{\varphi_T} qN_W\Omega_W \cong -2qN_W\Omega_W W_{-1}\left[-a\left(n_i/N_W\right)^2\right] \quad (22)$$

where $\Omega_W = AL_W$ is the characteristic well volume depending in particular on layout parameter of the anode-cathode spacing. The temperature dependence of $Q_C$ enters into this model only through the temperature dependence of the intrinsic silicon concentration and turns out to be noticeably stronger than the temperature dependence of $V_H$. The critical charge is proportional here to the characteristic value, close in meaning to the technological parameter called the Gummel number $Q_w \equiv qN_w\Omega_w$ which is important in bipolar transistors [26].

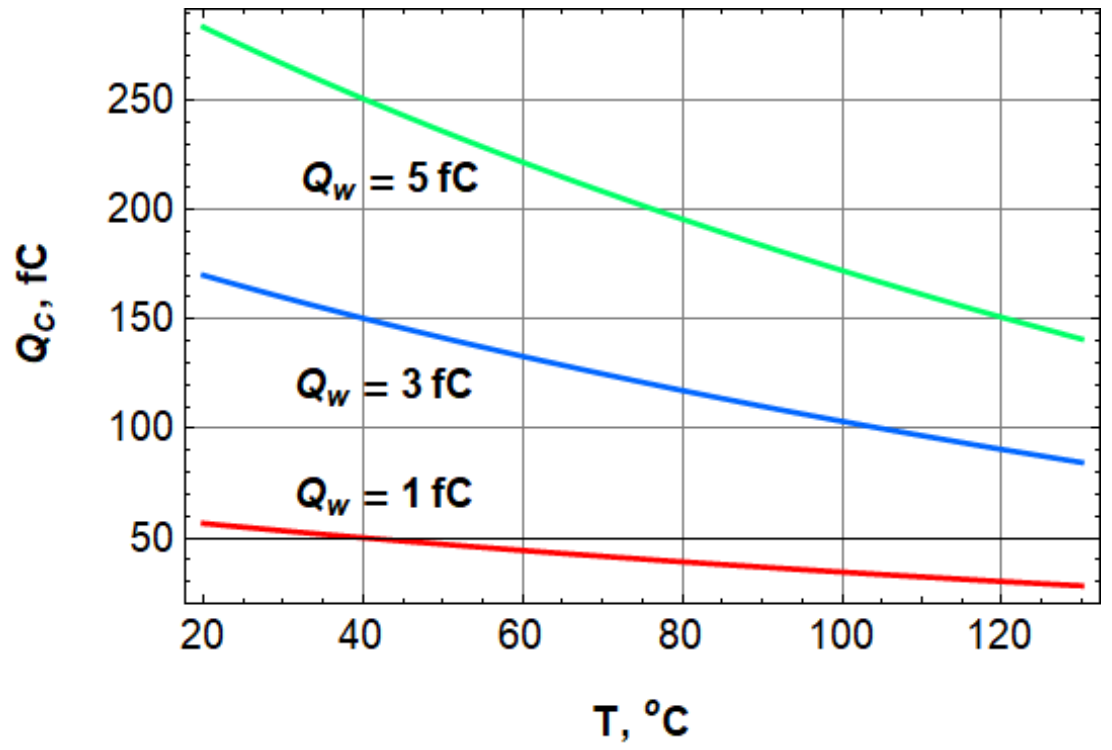

Fig. 7 Critical charge simulated with (22) as functions of temperature at different Gummel charges and $N_w = 10^{16}$ cm$^{-3}$.

Increasing $Q_w$, for example by increasing doping or the well effective sizes, leads to a decrease in $Q_C$ (see Fig.7) and ultimately to lowering of the IC sensitivity to the SELs.

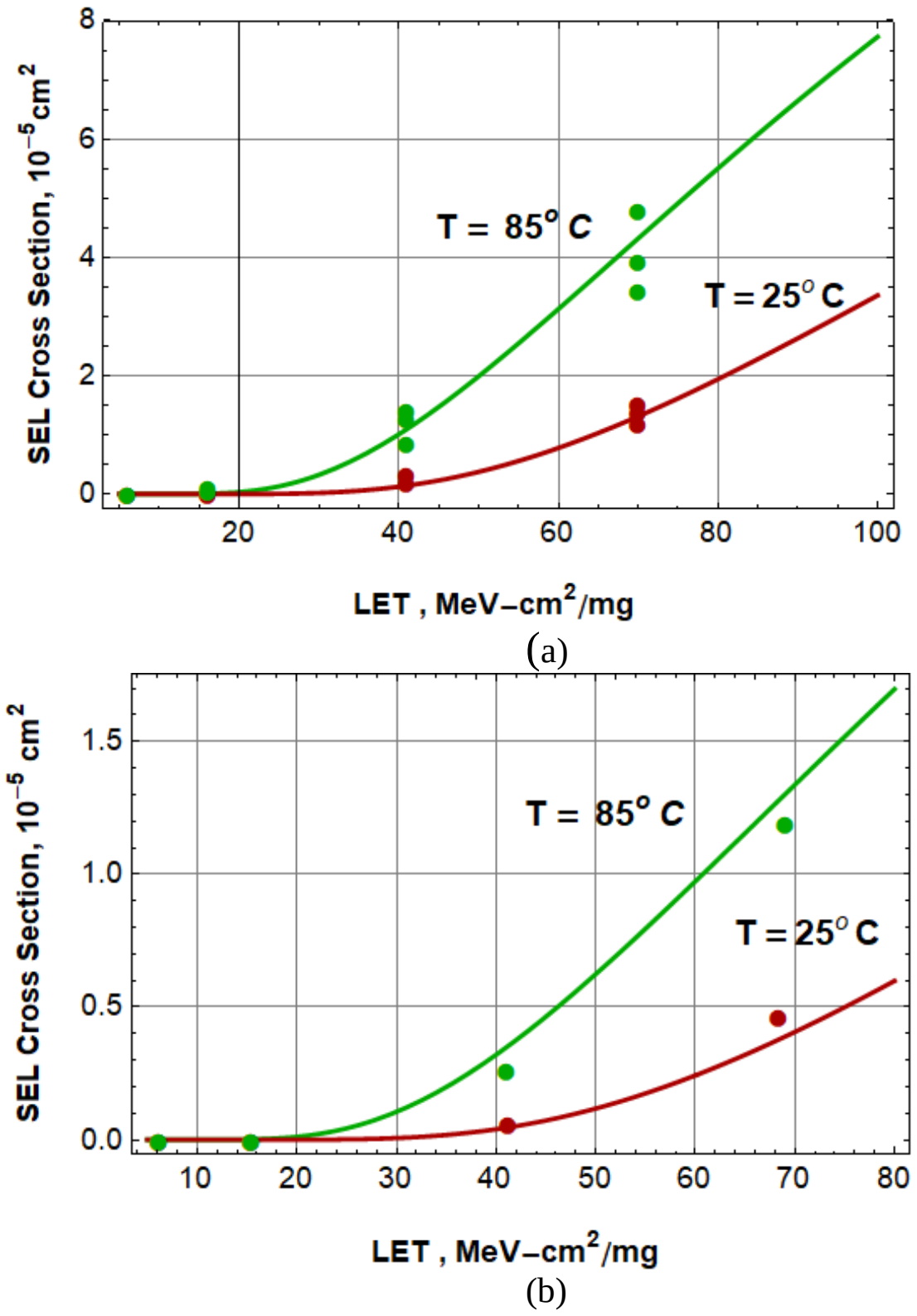

Fig. 8 Validation of the SEL cross section model at different temperatures

An example of model validation based on our own experimental data is shown in Fig. 8 for *type D* (a) and *type E* (b) devices (see description in [21]). The comparison of measured and simulated temperature dependencies of cross section measured at $\Lambda$ = 69 MeV-cm$^2$/mg for a *device A* is shown in Fig. 9.

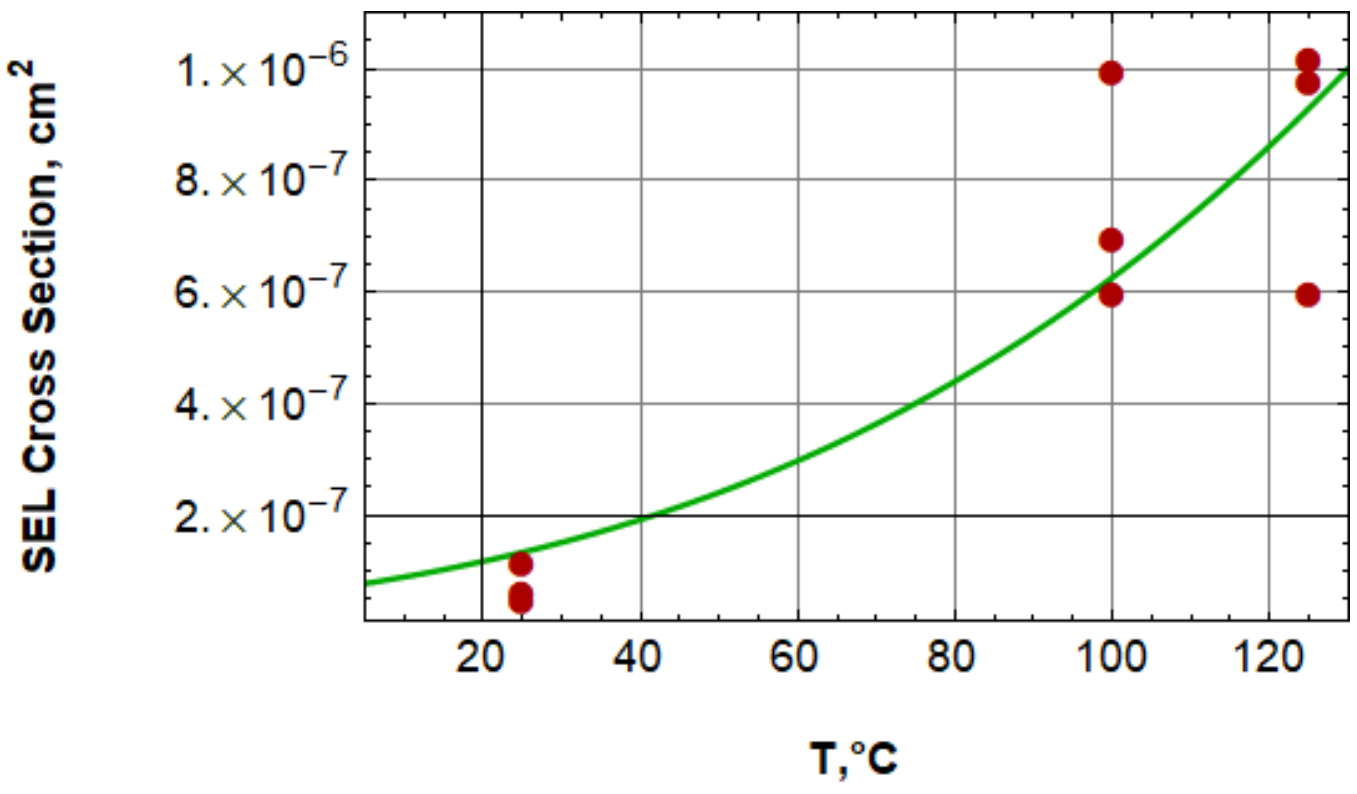

Fig. 9 Validation of the SEL cross section model at different measurement temperatures ( $Q_w = 5.4$ fC, $N_w = 6\times10^{16}$ cm$^{-3}$, A =1.2×10$^{-5}$ cm$^2$ )

Different points in Figs. 8-9 correspond to different samples under test. Apart from the well doping $N_W$, the model has only one free fitting constant $Q_W/t_{eff}$. Here the value of $t_{eff}$ was assumed to be fixed $t_{eff}$ =100 nm, and "the Gummel charge" turned out to be in the range $Q_W \cong$ 1-5 fC for all simulated devices.